\documentclass[journal=jacsat,manuscript=article, BackFigs]{achemso}

\usepackage{chemformula} 
\usepackage{graphicx}
\usepackage{epstopdf}
\usepackage{bm,amsmath,amssymb} 
\usepackage{color, soul} 
\usepackage{dcolumn}   
\usepackage{multirow}
\usepackage{makecell}
\usepackage{booktabs}

\usepackage{hyperref}

\newcommand{\eg}{{\em e.\,g.}}
\newcommand{\note}[1]{{\color{black}{{#1}}}}

\newcommand{\OP}{O$^{p}$}
\newcommand{\ONP}{O$^{np}$}
\author{Qisheng Yu}
\affiliation{Key Laboratory for Quantum Materials of Zhejiang Province, Department of Physics, School of Science and Research Center for Industries of the Future, Westlake University, Hangzhou, Zhejiang 310030, China}
\altaffiliation{Contributed equally to this work}
\author{Jiawei Huang}
\affiliation{Key Laboratory for Quantum Materials of Zhejiang Province, Department of Physics, School of Science and Research Center for Industries of the Future, Westlake University, Hangzhou, Zhejiang 310030, China}
\altaffiliation{Contributed equally to this work}
\author{Changming Ke}
\affiliation{Key Laboratory for Quantum Materials of Zhejiang Province, Department of Physics, School of Science and Research Center for Industries of the Future, Westlake University, Hangzhou, Zhejiang 310030, China}
\alsoaffiliation{Institute of Natural Sciences, Westlake Institute for Advanced Study, Hangzhou, Zhejiang 310024, China}
\author{Zhuang Qian}
\affiliation{Key Laboratory for Quantum Materials of Zhejiang Province, Department of Physics, School of Science and Research Center for Industries of the Future, Westlake University, Hangzhou, Zhejiang 310030, China}
\author{Liyang Ma}
\affiliation{Key Laboratory for Quantum Materials of Zhejiang Province, Department of Physics, School of Science and Research Center for Industries of the Future, Westlake University, Hangzhou, Zhejiang 310030, China}
\author{Shi Liu}
\email{liushi@westlake.edu.cn}
\affiliation{Key Laboratory for Quantum Materials of Zhejiang Province, Department of Physics, School of Science and Research Center for Industries of the Future, Westlake University, Hangzhou, Zhejiang 310030, China}
\alsoaffiliation{Institute of Natural Sciences, Westlake Institute for Advanced Study, Hangzhou, Zhejiang 310024, China}

\title{Semiconducting nonperovskite ferroelectric  oxynitride designed {\em ab~initio}}%

\begin{document}

\newpage
\begin{abstract}
Recent discovery of HfO$_2$-based and nitride-based ferroelectrics that are compatible to the semiconductor manufacturing process have revitalized the field of ferroelectric-based nanoelectronics. Guided by a simple design principle of charge compensation and density functional theory calculations, we discover HfO$_2$-like mixed-anion materials, TaON and NbON, can crystallize in the polar $Pca2_1$ phase with a strong thermodynamic driving force to adopt anion ordering spontaneously. Both oxynitrides possess large remnant polarization, low switching barriers, and unconventional negative piezoelectric effect, making them promising  piezoelectrics and ferroelectrics. Distinct from HfO$_2$ that has a wide band gap, both TaON and NbON can absorb visible light and have high charge carrier mobilities, suitable for ferroelectric photovoltaic and photocatalytic applications. This new class of multifunctional nonperovskite oxynitride containing economical and environmentally benign elements offer a platform to design and optimize high-performing ferroelectric semiconductors for integrated systems.

\end{abstract}
\newpage
The discovery of ferroelectricity in HfO$_2$-based thin films~\cite{Boscke11p102903} is revolutionizing the research and development for ferroelectric-based electronic devices. The demonstrated nanoscale ferroelectricity~\cite{Lee20p1343} and excellent compatibility with the modern complementary metal oxide semiconductor (CMOS) technology make HfO$_2$-based ferroelectrics the leading candidate material for next-generation nonvolatile information storage technology~\cite{Luo20p1391,Kim21peabe1341}. The realization of ferroelectricity in this bindary oxide also offers several conceptual breakthroughs. First, despite the admittedly importance of perovskite ferroelectrics that often contain 3$d$ elements such as Ti, nonperovskite oxides containing 5$d$ elements like Hf could afford great potential for new silicon-compatible ferroelectrics~\cite{Luo20p1391}. The finding that the polar orthorhombic phase (PO) of HfO$_2$ (space group $Pca2_1$) responsible for the ferroelectricity in thin films is actually higher in energy than the bulk monoclinic (M) phase (space group $P2_1/c$)~\cite{Huan14p064111,Sang15p162905} implies the importance of exploring metastable ferroelectric materials. 

Another progress in the search of ferroelectrics for semiconductor process integration is the successful synthesis of nitride ferroelectrics represented by nitride perovskites~\cite{Talley21p1488}  and doped III-V piezoelectrics~\cite{Fichtner19p114103,Hayden21p044412}. 
Though many nitride perovskites have long been predicted to be thermodynamically stable based on first-principles density functional theory (DFT) calculations~\cite{Fang17p014111, Sun19p732}, it was only until very recently that high-quality polycrystalline thin films of pure nitride perovskite, lanthanum tungsten nitride (LaWN$_3$), was synthesized with physical vapor deposition~\cite{Talley21p1488}. Previous DFT calculations predicted a switchable polarization of 61 $\mu$C/cm$^2$ in LaWN$_3$~\cite{Fang17p014111}. The synthesized LaWN$_3$ indeed has a polar symmetry and exhibits a piezoelectric response stronger than other known nitrides. Different from the strategy of stabilizing the polar phase in HfO$_2$-based ferroelectrics and LaWN$_3$ by fine tuning synthesis parameters, the occurrence of ferroelectricity in AlN-based ferroelectrics is to destabilize AlN that is already in a polar phase~\cite{Fichtner19p114103, Hayden21p044412}. For example, by doping Sc into AlN, a switchable polarization (80--110 $\mu$C/cm$^2$) has been realized in Al$_{1-x}$Sc$_x$N~\cite{Fichtner19p114103}. Taking advantage of decades of advances in nitride semiconductor technology, it is envisioned that the integration of nitride ferroelectrics with the mainstream semiconductor industry could enable novel device types for widespread applications such as nonvolatile memory, neuromorphic electronics, and negative capacitance transistors~\cite{Hong21p1445}. 

Both hafnia-based and nitride-based ferroelectrics open up opportunities for incorporating ferroelectric functionalities into integrated circuits. However, there are few of them. This leads us to the question of how to obtain more CMOS-compatible ferroelectrics. \note{DFT-based materials modeling offers a great platform to discover and design new ferroelectric materials with desired properties. Recent theoretical studies have predicted a few oxynitride-based ferroelectric materials, such as Ca$_3$Nb$_2$N$_2$O$_5$~\cite{Gou20p2815}, YGeO$_2$N, LaSiO$_2$N, and LaGeO$_2$N~\cite{Cohen21pJ56}. The combination of predictive high-throughput computations and the tools of data science further speeds up the material discovery~\cite{Palummo23p1548,Yu20p45023,Bouri18p2771}. After a promising material is predicted in theory, an immediate question is how to synthesize it in experiments; it often took months if not years of efforts to figure out the synthesizing protocol for a completely new compound.} In this work, \note{we aim to design new ferroelectrics that have structures similar to the polar phase of HfO$_2$ based on experimentally synthesized compounds.} Guided by a simple design principle of charge compensation and using the experimentally synthesized ferroelectric phase of HfO$_2$ as the template, we discover that the nonpolar monoclinic TaON and NbON that have already been synthesized can crystallize in the polar $Pca2_1$  phase that is dynamically stable, as supported by phonon spectrum calculations and {\em ab initio} molecular dynamic (AIMD) simulations. Based on first-principles DFT calculations, we demonstrate that $Pca2_1$ TaON and NbON have low switching barriers of $\approx$0.25 eV, large spontaneous polarization of $\approx$50 $\mu$C/cm$^2$, and direct band gaps of $\approx$2 eV, making them intrinsic semiconducting ferroelectrics with nontoxic elements. Owing to the dispersed 4$d$ and 5$d$ orbitals, both oxynitrides acquire light electron and hole effective masses and high carrier mobility of $\approx$400 cm$^2$V$^{-1}$s$^{-1}$, competing favorably with superstar halide perovskites such as CH$_3$NH$_3$PbI$_3$~\cite{Dong15p967,Brenner15p4754}. These highly polarized and visible photoresponsive oxynitrides have great technological potential for use in ferroelectric photovoltaics~\cite{Butler15p838,Bennett08p17409,Huang19p14520, Shi19p18334}, multienergy harvester, and photocatalyst applications.

We use the ferroelectric $Pca2_1$ phase of HfO$_2$ as the template structure. In a simple ionic picture, after replacing all Hf$^{4+}$ cations with Ta$^{5+}$, an element sitting right next to hafnium in the periodic table, we need to replace half of the anions (O$^{2-}$) with more-negatively-charged anions such as N$^{3-}$ in order to maintain the charge neutrality. This leads to a ternary mixed-anion oxynitride, TaON. \note{Similarly, NbON is chosen as another candidate as TaON and NbON possess the same polymorphs.} 
Before performing any calculations, we present a few evidences justifying this simple design principle.  First, the bulk phase of HfO$_2$ adopts the $P2_1/c$ monoclinic phase,  while TaON and NbON in the same space group have already been synthesized in experiments and extensively studied because of high photocatalytic activity under visible light~\cite{Hitoki02p1698,Yashima07p588,Kikuchi15p711}.
It is thus worthy to explore whether TaON and NbON could crystallize in the polar $Pca2_1$ phase, similar to that in HfO$_2$. Additionally, the $Pca2_1$ phase of HfO$_2$ contains two symmetry-inequivalent oxygen atoms: a three-fold coordinated oxygen atom that has polar local displacement (\OP) and a fourfold-coordinated nonpolar oxygen (\ONP), as illustrated in Fig.~1a. Since N$^{3-}$ anions are more negatively charged than O$^{2-}$, they expectedly will occupy fourfold-coordinated \ONP~sites, leading to a spontaneous ordering of N$^{3-}$ anions without affecting polar oxygen atoms. \note{
We computed the energy difference between a unit cell of $Pca2_1$ TaO$_{1.75}$N$_{0.25}$ (NbO$_{1.75}$N$_{0.25}$) that has one N atom occupying the O$^p$ site and another configuration with one N atom occupying the O$^{np}$ site. We found that the O$^{np}$-site N substitution is 0.95 eV lower in energy than the O$^p$-site N substitution for TaO$_{1.75}$N$_{0.25}$, and 2.04 eV for NbO$_{1.75}$N$_{0.25}$. This confirms that N atoms have a strong tendency to replace the $np$-site O atoms.} Such anion ordering in monoclinic TaON has been suggested based on neutron and synchrotron powder diffraction and DFT studies~\cite{Yashima07p588}. 

First-principles DFT calculations are performed using \texttt{QUANTUM ESPRESSO} ~\cite{Giannozzi09p395502, Giannozzi17p465901} package with Garrity-Bennett-Rabe-Vanderbilt (GBRV) ultrasoft pseudopotentials~\cite{Garrity14p446}. The exchange-correlation function is Perdew-Burke-Ernzerhof (PBE) functional of generalized gradient approximation. The lattice constants and atomic positions are optimized with a plane-wave cutoff of 50 Ry and a charge density cutoff of 250 Ry. A $4 \times 4 \times 4$ $k$-point grid centered on $\Gamma$ point is used for Brillouin zone sampling. The total energy and force convergence threshold is set to 10$^{-5}$ Ry and 10$^{-4}$ Ry/Bohr, respectively. The phonon spectra are calculated using the finite displacement method implemented in \texttt{PHONOPY}~\cite{Togo15p1} code in conjunction with \texttt{QUANTUM ESPRESSO}. The structural stability at finite temperatures is gauged by $ab~initio$ molecular dynamics (AIMD) simulations implemented in Vienna $ab~initio$ simulation package (VASP)~\cite{Kresse96p11169, Kresse96p15}. We construct a $3\times3\times3$ supercell for AIMD simulations using a $\Gamma$-point sampling, an energy cutoff of 350 eV and an energy convergence threshold of 10$^{-5}$~eV. The Nos\'e-Hoover thermostat is applied for temperature controlling. The minimum energy paths (MEPs) of polarization reversal are determined using the  variable-cell nudged elastic band (NEB) technique implemented in \texttt{USPEX} code~\cite{Oganov06p244704,Lyakhov13p1172,Oganov11p227}. Noted that VCNBE allows strain relaxation during the polarization switching process~\cite{Huang22p144106}. The MEP is considered to reach convergence when the root-mean-square forces are lower than 0.03 eV/\AA~on images or the energy barrier remains unchanged for successive 10 steps. The carrier mobility is estimated using semiclassical Boltzmann transport using \texttt{Boltztrap}~\cite{Madsen06p67} within the constant relaxation time ($\tau$=0.2~ps) approximation,.

Different from the oxygen atoms in perovksite oxides that often possess similar local chemical environments, the \OP~ and \ONP~sites have distinct local bonding characteristics. 
We investigate the effect of anion ordering on the relative thermodynamic stability of different atomic configurations of $Pca2_1$ TaON. For a 12-atom unit cell, there are five possible configurations, N$^{np}$, N$^p$, N$^{diag}$, N$^{plane}$, and N$^{tetra}$, that are named based on the ordering of N atoms (Fig.~1b-f). As expected, the N atom has a strong tendency to occupy the fourfold-coordinated \ONP~site. The $Pca2_1$-N$^{np}$ configuration that has all N atoms occupying \ONP~sites is the most stable, followed by the $Pca2_1$-N$^{p}$ configuration ($\Delta E$= 28 meV/atom) that has N atoms located at \OP~sites, whereas all the other configurations are much higher in energy (Table~\ref{energy}). For example, $Pca2_1$-N$^{plane}$ is unstable such that it spontaneously evolves to the $P2_1/c$ phase with all N atoms being fourfold-coordinated to tantalum
after the structural optimization. Such strong tendency of O/N anion ordering is beneficial for the emergence of ferroelectricity as the polar oxygen atom in $Pca2_1$-N$^{np}$ TaON acquires a similar local environment to that in ferroelectric HfO$_2$. The spontaneous O/N anion ordering also presents in NbON (Table~\ref{energy}). We note that the monoclinic $P2_1/c$-N$^{np}$ configuration remains the most stable thermodynamically for TaON and NbON. As we discussed above, an important lesson learned from the discovery of ferroelectric HfO$_2$ is that stabilizing a high-energy metastable phase in thin films is feasible experimentally. Therefore, we argue that it is \note{worthwhile to attempt synthesizing } $Pca2_1$-N$^{np}$ TaON and NbON (if they are dynamically stable, see discussions below).

For HfO$_2$, the antiferroelectric-like $Pbca$ phase is the second most stable~\cite{Huan14p064111} and is 4~meV/atom lower in energy than $Pca2_1$~\cite{Liu19p054404}. 
In comparison, the $Pbca$-N$^{np}$ and $Pca2_1$-N$^{np}$ phases in TaON have comparable energies (difference within 1 meV/atom). The $Pca2_1$-N$^{np}$ phase of NbON becomes more stable than $Pbca$-N$^{np}$. This is probably helpful for the stabilization of the metastable $Pca2_1$-N$^{np}$ phase in NbON. \note{Hereafter, unless otherwise specified, all results are for configuration $Pca2_1$-N$^{np}$.}

The structural stability of the $Pca2_1$ phase of TaON and NbON is examined by computing the phonon spectrum. A (meta)stable material situating at a local minimum of the potential energy surface 
will have all positive phonon frequencies. The phonon spectra of $Pca2_1$ TaON and NbON reveal no imaginary frequencies in the whole Brillouin zone (Fig.~\ref{phonon}a-b), confirming the dynamical stability of both compounds. \note{This is similar to HfO$_2$ where several metastable phases such as $P2_1/c$, $Pbca$, $P4_2/nmc$ that are higher in energy than the $P2_1/c$ phase are all dynamically stable as confirmed by phonon spectra exhibiting no imaginary modes~\cite{Fan22p32}.} We further perform AIMD simulations at elevated temperatures to check whether both compounds will be stable against larger atomic distortions due to thermal fluctuations. Figure \ref{phonon}c plots the fluctuations of the total energy for $Pca2_1$ TaON (NbON) at 400, 600, and 800~K during AIMD simulations, showing no sign of structural destruction or reconstruction to other phases. This serves as a strong evidence supporting the room-temperature stability of $Pca2_1$ TaON and NbON. Additional AIMD simulations reveal that TaON becomes nonpolar when the temperature is above 2000~K. We analyze the distribution of local displacements ($\Delta z$) of oxygen atoms along the polar axis relative to the nearest Hf atomic plane using configurations sampled from a 10 ps equilibrium trajectory. With increasing temperature, the peak position shifts toward lower values, indicating the polar oxygen atoms move closer to their nearest Hf plan es. 

The switching barriers are computed using the VCNEB method that takes into account the strain relaxation effect during the switching~\cite{Huang22p144106}. There are two possible switching pathways in ferroelectric HfO$_2$, the shift-inside (SI) pathway that has \OP~atoms moving inside two Hf atomic planes (Fig.~\ref{neb}a) and the shift-across (SA) pathway that has \OP~atoms moving across the Hf atomic plane (Fig.~\ref{neb}b)~\cite{Wei22p154101,Choe21p8}. Consistent with previous studies~\cite{Wei22p154101}, the SI barrier in HfO$_2$ is 0.38 eV, much lower than the SA barrier of 0.80 eV. Interestingly, the opposite trend is found in TaON and NbON: the SA pathway becomes kinetically favored over the SI pathway. This is consistent with AIMD simulations in which polar oxygen atoms tend to move toward their nearest Hf atomic planes with increasing temperatures. Importantly, both oxynitrides are switchable as the SA barrier is $\approx$0.25 eV, comparable to the switching barrier of prototypical ferroelectric like PbTiO$_3$ (0.17 eV)~\cite{Huang22p144106}. These results demonstrate the presence of switchable polarization in $Pca2_1$ TaON and NbON, a hallmark feature of ferroelectricity.

We determine the magnitude of the spontaneous polarization with the Berry-phase approach~\cite{KingSmith93p1651, Spaldin12p2} by tracking the change in Berry phase during the SA pathway. As shown in Fig.~\ref{neb}c, both TaON and NbON have large and nearly identical magnitudes of polarization, $\approx$50 $\mu$C/m$^2$. 
The Born effective charge tensors ($Z$) of transition metal, oxygen, and nitrogen atoms are reported in Table~\ref{para}. An interesting observation is that for the $Z_{33}$ component, transition metal ($M$) and nitrogen atoms have effective charges ($\approx$$+$6 for $M$ and $\approx$$-$4 for N) larger than their nominal ionic valence ($+5$ for $M$ and $-3$ for N), while the O ion acquires a value of $\approx-2.1$, close to the nominal value. In comparison, the $Z_{33}$ value of O in perovskite ferroelectric BaTiO$_3$ is anomalously large ($-5.7$)~\cite{Zhong94p3618}, a signature of strong charge transfer and covalency. We suggest that the $M$-N bond is likely more covalent that the $M$-O bond along the polar axis. 

Additionally, the SA pathway is associated with an unconventional polarization-strain coupling that has the strain along the polar axis ($\eta_c$) increasing with reducing polarization, hinting at negative longitudinal piezoelectric effect~\cite{Liu17p207601}. Indeed, TaON has $e_{33}$ of $-1.70$ C/m$^2$ and $d_{33}$ of $-7.63$ pm/V, while NbON has $e_{33}$ of $-0.85$ C/m$^2$ and $d_{33}$ of $-5.92$ pm/V, all possessing a negative sign and a decent magnitude that compares well to those of commercial piezoelectrics such as Al$_{0.92}$Sc$_{0.08}$N ($\approx$10 pm/V) and LiNbO$_3$ ($\approx$10 pm/V)~\cite{Talley21p1488}. \note{The piezoelectric coefficient can be decomposed into the clamped-ion contribution (fixed ion with vary strains) and the internal-strain contribution arising from ion relaxations~\cite{Szabo98p4321}. Our calculations suggest that the clamped-ion ($-0.17$ C/m$^2$ for NbON and $-1.24$ C/m$^2$ for TaON) and internal-strain contributions ($-0.48$ C/m$^2$ for NbON and $-0.62$ C/m$^2$ for TaON) are both negative, thus leading to a macroscopic negative piezoelectric response. 
By combining piezoelectrics possessing negative piezoelectric effect with conventional piezoelectrics, it is possible to achieve on-demand design of $e_{33}$ and precise control of the electromechanical response~\cite{Liu17p207601}.}

The PBE band structures of TaON and NbON in the $Pca2_1$ phase are
shown in Fig.~\ref{ele}a-b, revealing a direct band gap of 1.86 and 1.32~eV, respectively, at the X point. The projected density of states (PDOS) indicate that bands near the valence band maximum (VBM) take mostly  N-$2p$ character with some contributions from O-$2p$ and Ta-$5d$ (Nb-$4d$) characters while those near the conduction band minimum (CBM) mainly consist of $d$-orbitals of the transition metal. We estimate the effective masses of electrons ($e$) and holes ($h$) based on the dispersion relationship along X-S with results reported in Table~\ref{para}. Owing to the spatially more dispersed $5d$ ($4d$) orbitals, the effective electron and hole masses are almost identically light in ferroelectric TaON (NbON), both at a value of $\approx$0.4. This resembles to that in tetragonal CH$_3$NH$_3$PbI$_3$~\cite{Frohna18p1829} and is beneficial for balanced transport between the electrons and holes. The theoretical spatially averaged charge carrier mobility, $\mu$, is plotted in Fig.~\ref{ele}c as a function of charge carrier concentration ($\rho_e$ and $\rho_h$). At a low carrier concentration ($\rho<10^{19}$/cm$^3$), the electron and hole mobilities ($\mu_e$ and $\mu_h$) are insensitive to $n$.  Notably, we predict $\mu_h$ of 455 and $\mu_e$ of 332~cm$^2$V$^{-1}$s$^{-1}$ in TaON, and $\mu_h$ of 464 and $\mu_e$ of 301~cm$^2$V$^{-1}$s$^{-1}$ in NbON. These values are at least one order of magnitude higher than those ($<10$~cm$^2$V$^{-1}$s$^{-1}$) in CH$_3$NH$_3$PbI$_3$ estimated with the same method that assumes a longer carrier relaxation time of 1~ps~\cite{Motta15p12746}. 

It is well known that PBE underestimates the band gap due to the  the remnant self-interaction error in the approximation to the exchange-correlation functional. To address this issue, we further compute the band structures using the newly developed pseudohybrid Hubbard density functional, extend Agapito–Cuetarolo–Buongiorno Nardelli (eACBN0)~\cite{Agapito13p165127, Lee20p043410, Tancogne-Dejean20p155117}. The eACBN0 function is essentially a DFT+$U$+$V$ method with self-consistently computed  Hubbard $U$ ($V$) parameters that account for the onsite (intersite) Coulomb interaction. Particularly for covalent semiconductors (\eg, Si and GaAs), the band gaps predicted with eACBN0 are much better than PBE values, and are comparable with advanced methods such as the Heyd-Scuseria-Ernzerhof (HSE) hybrid density functional and
$GW$ approximations but at a PBE-level computational cost~\cite{Huang20p165157, Lee20p043410, Yang22p195159}. As shown in Fig.~\ref{ele}a-b,  eACBN0 predicts a larger band gap of 2.31 eV for TaON and 1.75 eV for NbON than PBE, while both functionals predict similar band dispersion relationships (and thus effective masses and carrier mobilities). Therefore, TaON and NbON are indeed intrinsic ferroelectric semiconductors suited to absorb visible light. \note{Additionally, we found that the structural parameters obtained with PBE and eACBN0 are comparable. Interestingly, PBE and eACBN0 predict nearly identical $e_{33}$ for NaON, while eACBN0 predicts a lower magnitude of $e_{33}$ ($-0.80$ C/m$^2$) than PBE ($-1.70$ C/m$^2$) for TaON. }

The direct-band-gap nature of proposed oxynitrides makes them better solar absorbers than Si which is known to have poor absorption of low-energy photons below the direct band gap~\cite{Xiang13p118702}. We compare the PBE absorption spectra of TaON, NbON, Si, and CH$_3$NH$_3$PbI$_3$ in Fig.~\ref{ele}d. It is evident that both TaON and NbON possess superior absorption performances than Si in the visible region, and NbON competes favorably with CH$_3$NH$_3$PbI$_3$. Utilizing well-developed epitaxy techniques, TaON and NbON could be made as thin-film solar absorbers for efficient light absorption. \note{Compared with conventional perovskite ferroelectric materials whose band gaps are around 3 eV, the band gaps of TaON and NbON are relatively small and may be associated with larger leakage currents at high electric fields. We suggest that these two ferroelectric semiconductors are probably more suitable for photovoltaic-related applications.
Further studies are needed to resolve the leakage current issue.}

Finally, compared with defect-engineered semiconducting ferroelectric perovskite oxides such as (KNbO$_3$)$_{1-x}$(BaNi$_{0.5}$Nb$_{0.5}$O$_{3-\delta})_x$~\cite{Grinberg13p509}  and (Na$_{0.5}$Ba$_{0.5}$TiO$_3$)$_{1-x}$(BaTi$_{0.5}$Ni$_{0.5}$O$_{3-\delta})_x$~\cite{Xiao18p1805802}, TaON and NbON are intrinsic semiconducting photoferroics functioning in the visible range with much simpler compositions. Given that tantalum oxides/nitrides~\cite{Lima10p319,Zhu13p1151} and niobium oxides/nitrides~\cite{Matsui05pF54,Bower20p45444} have all demonstrated applications in the production of microelectronics, both oxynitrides are expected to have CMOS compatibility.

Using experimentally synthesized ferroelectric HfO$_2$ as the template, we design a new class of nonperovskite oxynitrides represented by TaON and NbON in the space group of $Pca2_1$ that have promising CMOS compatibility. Results from first-principle calculations demonstrate that both oxynitrides are multifunctional ferroelectric semiconductors, with key material parameters such as magnitude of spontaneous polarization, piezoelectric coefficient, carrier mobility, and light absorption being competitive with many state-of-the-art materials such as Si and halide perovskites. Combined with the enhanced electron-hole separation in ferroelectrics~\cite{Grinberg13p509,Frost14p2584,Liu15p693}, the visible-light-responsive oxynitrides with simple chemical composition and nontoxic elements could act as efficient components in photovoltaic and photocatalytic devices. The design strategy in this work also highlights the importance of nonperovskite structure and $4d$/$5d$ elements for the discovery and development of ferroelectric semiconductors for next-generation energy and information technology. \\

This work is supported by National Key R\&D Program of China (2021YFA1202100), National Natural Science Foundation of China (12074319), Natural Science Foundation of Zhejiang Province (2022XHSJJ006), and Westlake Education Foundation. The computational resource is provided by Westlake HPC Center. We acknowledge useful discussions with Dr. Fan Zheng.

\clearpage
\begin{table}[ht]
    \centering
    \caption{Energy ($\Delta E$ in meV/atom) and lattice constants (in \AA) of TaON and NbON computed with PBE. The energy of $Pca2_1$-N$^{np}$ is chosen as the reference. The polar axis in $Pca2_1$ is along the $c$-axis. As a comparison, the energy of $P2_1/c$ and $Pbca$ of HfO$_2$ is $-28$ and $-4$ meV/atom, respectively, relative to $Pca2_1$. \note{Experimental lattice parameters for the $P2_1/c$-N$^{np}$ configuration are reported in the parentheses~\cite{Weishaupt77p261,Fang01p1248}}.}
    \label{energy}
    \begin{tabular}{ccccccc}
        \hline
        \hline
        \specialrule{0em}{1pt}{1pt}
        {TaON} & $Pca2_1$-N$^{np}$ & $Pca2_1$-N$^{p}$ & $Pca2_1$-N$^{diag}$ & $Pca2_1$-N$^{tetra}$ & $P2_1/c$-N$^{np}$ & $Pbca$-N$^{np}$\\
        \hline
        \specialrule{0em}{1pt}{1pt}
        $\Delta E$ & 0 & 28 & 54 & 91 & $-63$ & $-1$\\ 
        $a$ & 5.16 & 5.81 & 5.28 & 5.18 & 5.23 \note{(5.18)} & 5.14\\
        $b$ & 4.91 & 4.80 & 5.12 & 4.92 & 5.01 \note{(4.96)} & 9.78\\
        $c$ & 5.03 & 5.11 & 5.01 & 5.01 & 5.07 \note{(5.03)} & 5.06\\
        $\alpha$ & 90 & 90 & 90 & 90 & 90 \note{(90)} & 90\\
        $\beta$ & 90 & 90 & 99.8 & 90 & 90 \note{(90)} & 90\\
        $\gamma$ & 90 & 90 & 90 & 89.0 & 99.7 \note{(99.6)} & 90\\
        \specialrule{0em}{1pt}{1pt}
        \hline
        \hline
        \specialrule{0em}{1pt}{1pt}
        {NbON} & $Pca2_1$-N$^{np}$ & $Pca2_1$-N$^{p}$ & $Pca2_1$-N$^{diag}$ & $Pca2_1$-N$^{tetra}$ & $P2_1/c$-N$^{np}$ & $Pbca$-N$^{np}$\\
        \hline
        \specialrule{0em}{1pt}{1pt}
        $\Delta E$ & 0 & unstable & 48 & 92 & $-52$ & 5\\ 
        $a$ & 5.19 & - & 5.28 & 5.18 & 5.22 \note{(5.19)} & 5.14\\
        $b$ & 4.90 & - & 5.11 & 4.88 & 5.00 \note{(4.97)} & 9.74\\
        $c$ & 5.02 & - & 5.01 & 5.03 & 5.06 \note{(5.03)} & 5.05\\
        $\alpha$ & 90 & - & 90 & 90 & 90 \note{(90)} & 90\\
        $\beta$ & 90 & - & 100.0 & 90 & 90 \note{(90)} & 90\\
        $\gamma$ & 90 & - & 90 & 90.7 & 99.8 \note{(100.2)} & 90\\
        \specialrule{0em}{1pt}{1pt}
        \hline
        \hline
    \end{tabular}
\end{table}
\clearpage
\newpage
\begin{table}[htp]
  \centering
  \caption{Spontaneous polarization along the $c$-axis ($P_s$ in $\mu$C/cm$^2$), piezoelectric coefficients ($d_{33}$ in pm/V and $e_{33}$ in C/m$^2$), electron and hole effective masses ($m_e^*$ and $m_h^*$ in electron mass $m_0$), band gaps computed with PBE and eACBN0 ($E_g^{\rm PBE}$ and $E_g^{\rm eACBN0}$ in eV), and Born effective charge tensors ($Z$) for O, N, and transition metal ($M$) in ferroelectric $Pca2_1$-N$^{np}$ $M$ON ($M$=Ta, Nb).\label{para}}
  \begin{tabular}{cccccccccc}\hline\hline
  \multicolumn{1}{c|}{}   & \multicolumn{2}{c}{$P_s$}   
                          & $d_{33}$               & $e_{33}$
                          & $m^*_{e}$              & $m^*_{h}$    
                          & $E_g^{\rm PBE}$    & \multicolumn{2}{c}{$E_g^{\rm eACBN0}$}                
                          \\\hline
\multicolumn{1}{c|}{TaON} & \multicolumn{2}{c}{51.7}                   & $-$7.6                & $-$1.70            
                          & 0.435                      &  $-$0.414 
                          &  1.86                      & \multicolumn{2}{c}{ 2.31} 
                          \\
\multicolumn{1}{c|}{NbON} & \multicolumn{2}{c}{54.0}                   & $-$5.9                & $-$0.85           
                          & 0.425                   & $-$0.432 
                          &  1.32                      & \multicolumn{2}{c}{1.75} 
                          \\\hline\hline
  \multicolumn{1}{c|}{}   & \multicolumn{3}{c|}{$Z_{\rm N}$}                         
                          & \multicolumn{3}{c|}{$Z_{\rm O}$}
                          & \multicolumn{3}{c}{$Z_{M}$}           
                          \\\cline{2-10}
  \multicolumn{1}{c|}{}   & $Z_{11}$    & $Z_{22}$    & \multicolumn{1}{c|}{$Z_{33}$ }     
                          & $Z_{11}$    & $Z_{22}$    & \multicolumn{1}{c|}{$Z_{33}$ }           
                          & $Z_{11}$    & $Z_{22}$    & $Z_{33}$              
                          \\\hline
\multicolumn{1}{c|}{TaON} & $-$4.11       & $-$2.56       & \multicolumn{1}{c|}{$-$3.84}          
                          & $-$2.77       & $-$3.97       & \multicolumn{1}{c|}{$-$2.11}                 
                          & 6.87        & 6.53        & 5.94                   
                          \\
\multicolumn{1}{c|}{NbON} & $-$4.02       & $-$2.71       & \multicolumn{1}{c|}{$-$4.13}           
                          & $-$2.50       & $-$4.36       & \multicolumn{1}{c|}{$-$2.18}                  
                          & 6.51        & 7.07        & 6.30
                          \\\hline\hline                 
  \end{tabular}
\end{table}

\clearpage
\newpage
\begin{figure}[t]
\centering
\includegraphics[scale=0.8]{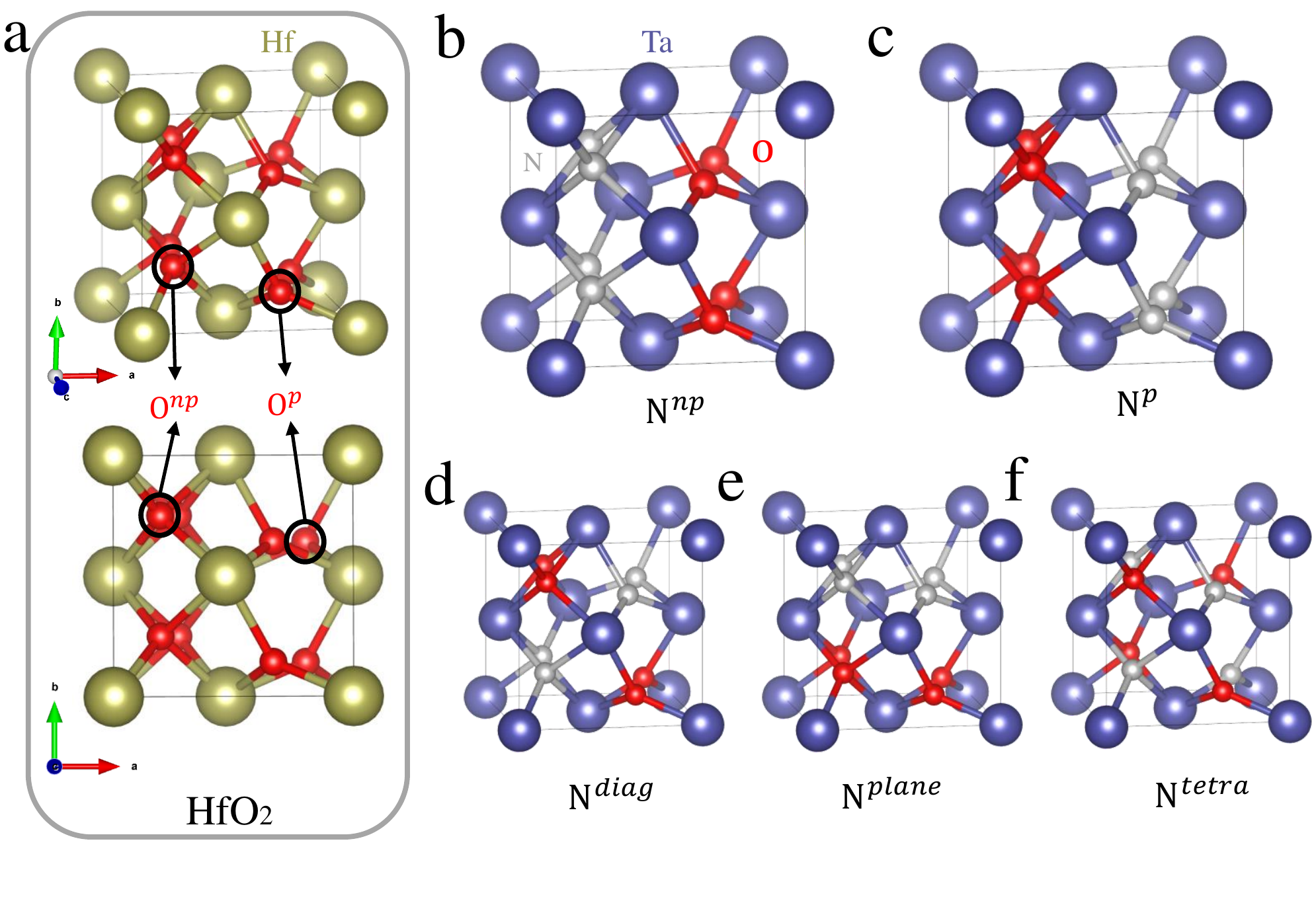}
\caption{(a) Crystal structure of the $Pca2_1$ phase of HfO$_2$ that has two symmetry-inequivalent oxygen
atoms, three-fold coordinated polar oxygen (\OP) and fourfold-coordinated nonpolar oxygen (\ONP). Illustrations of different types of anion ordering (b)-(f) in polar TaON named after the occupations of N atoms.}
\label{str}
\end{figure}

\clearpage
\newpage
\begin{figure}[t]
\centering
\includegraphics[scale=0.8]{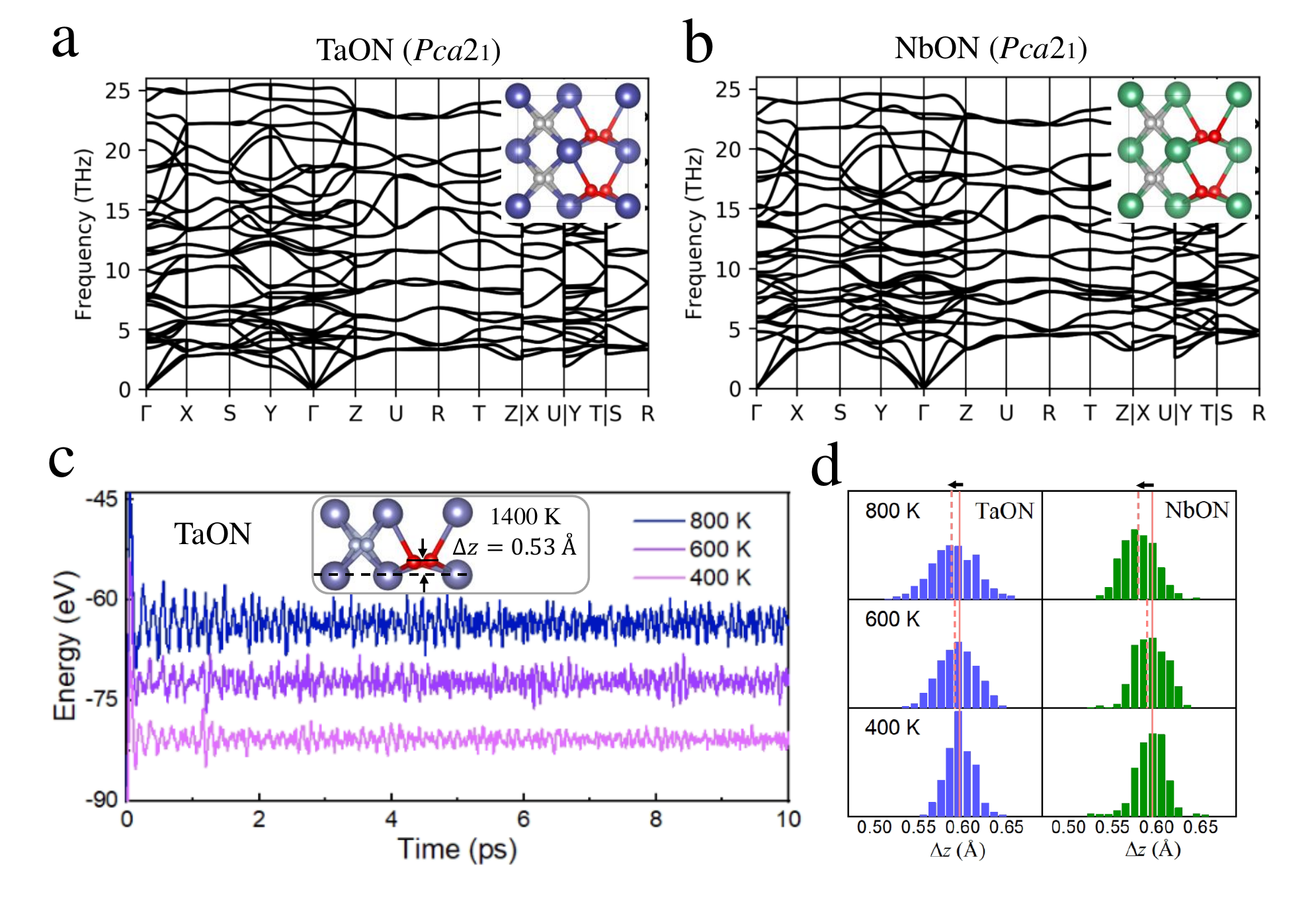}
\caption{Phonon dispersion relationships of (a) TaON and (b) NbON of \note{$Pca2_1$-N$^{np}$ configuration}. (c)  Energy evolution as a function of time in AIMD of TaON at 400, 600, 800~K. The inset shows the averaged unit cell at 1400~K. (d) Distribution of atomic displacements ($\Delta z$, see inset in c) of polar oxygen atoms relative to the nearest Hf atomic plane along the polar axis.}
\label{phonon}
\end{figure}

\clearpage
\newpage
\begin{figure}[t]
\centering
\includegraphics[scale=0.8]{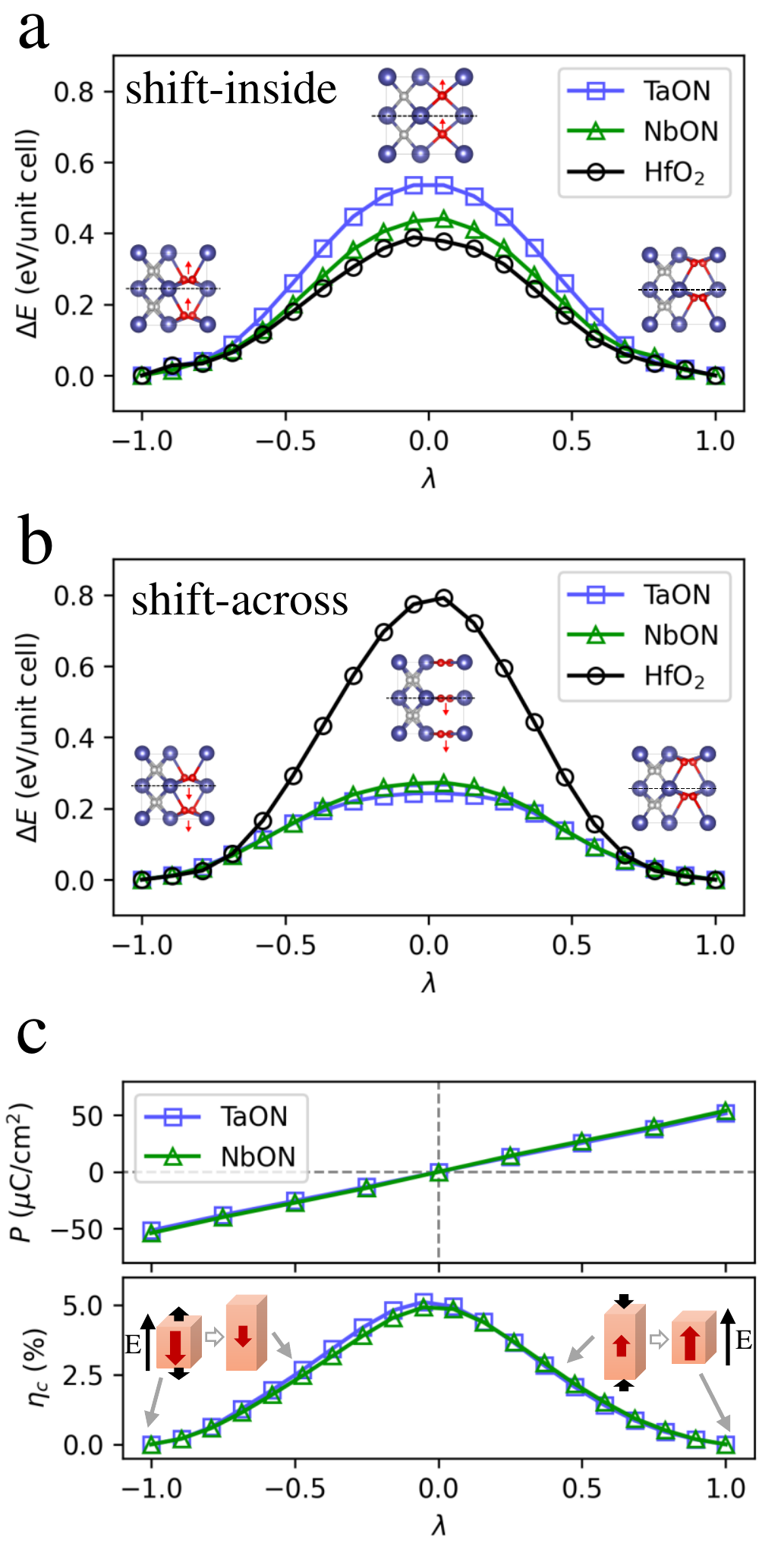}
\caption{Minimum energy paths of polarization reversal in ferroelectric $Pca2_1$-N$^{np}$ TaON and NbON, and HfO$_2$ identified with VCNEB for (a) shift-inside and (b) shift-across mechanisms. (c) Evolution of the polarization and the strain ($\eta_c$ defined as $c/c_0-1$) along the shift-across switching pathway. The inset illustrates the negative longitudinal piezoelectric effect.}
\label{neb}
\end{figure}

\clearpage
\newpage
\begin{figure}[t]
\centering
\includegraphics[scale=0.8]{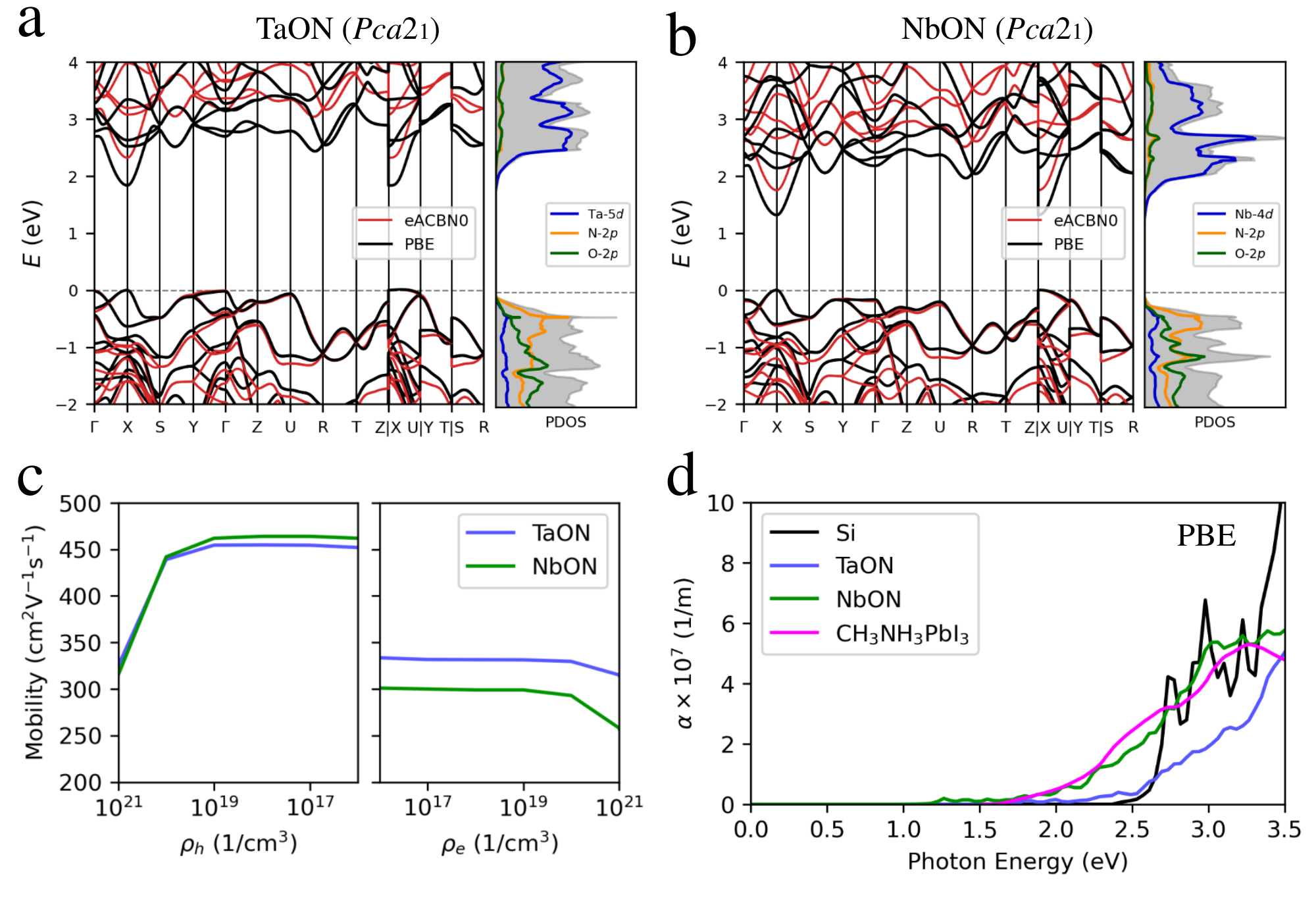}
\caption{Electronic properties of \note{$Pca2_1$-N$^{np}$} TaON and NbON. Band structures and projected density of states (PDOS) computed with PBE and eACBN0 for (a) TaON and (b) NbON. (c) Average hole and electron mobility for TaON and NbON as a function of carrier concentration ($\rho$). (d) Comparison of PBE absorption spectra of ferroelectric oxynitrides, Si, and CH$_3$NH$_3$PbI$_3$. At the PBE level, Si has an indrect band gap of 0.57~eV, and the absorption onset photon energy is at 2.4~eV.}
\label{ele}
\end{figure}

\bibliography{SL}

\providecommand{\latin}[1]{#1}
\makeatletter
\providecommand{\doi}
  {\begingroup\let\do\@makeother\dospecials
  \catcode`\{=1 \catcode`\}=2 \doi@aux}
\providecommand{\doi@aux}[1]{\endgroup\texttt{#1}}
\makeatother
\providecommand*\mcitethebibliography{\thebibliography}
\csname @ifundefined\endcsname{endmcitethebibliography}
  {\let\endmcitethebibliography\endthebibliography}{}
\begin{mcitethebibliography}{65}
\providecommand*\natexlab[1]{#1}
\providecommand*\mciteSetBstSublistMode[1]{}
\providecommand*\mciteSetBstMaxWidthForm[2]{}
\providecommand*\mciteBstWouldAddEndPuncttrue
  {\def\EndOfBibitem{\unskip.}}
\providecommand*\mciteBstWouldAddEndPunctfalse
  {\let\EndOfBibitem\relax}
\providecommand*\mciteSetBstMidEndSepPunct[3]{}
\providecommand*\mciteSetBstSublistLabelBeginEnd[3]{}
\providecommand*\EndOfBibitem{}
\mciteSetBstSublistMode{f}
\mciteSetBstMaxWidthForm{subitem}{(\alph{mcitesubitemcount})}
\mciteSetBstSublistLabelBeginEnd
  {\mcitemaxwidthsubitemform\space}
  {\relax}
  {\relax}

\bibitem[B\"{o}scke \latin{et~al.}(2011)B\"{o}scke, M\"{u}ller, Br\"{a}uhaus,
  Schr\"{o}der, and B\"{o}ttger]{Boscke11p102903}
B\"{o}scke,~T.~S.; M\"{u}ller,~J.; Br\"{a}uhaus,~D.; Schr\"{o}der,~U.;
  B\"{o}ttger,~U. Ferroelectricity in Hafnium Oxide Thin Films. \emph{Appl.
  Phys. Lett.} \textbf{2011}, \emph{99}, 102903\relax
\mciteBstWouldAddEndPuncttrue
\mciteSetBstMidEndSepPunct{\mcitedefaultmidpunct}
{\mcitedefaultendpunct}{\mcitedefaultseppunct}\relax
\EndOfBibitem
\bibitem[Lee \latin{et~al.}(2020)Lee, Lee, Lee, Jo, Yang, Kim, Chae, Waghmare,
  and Lee]{Lee20p1343}
Lee,~H.-J.; Lee,~M.; Lee,~K.; Jo,~J.; Yang,~H.; Kim,~Y.; Chae,~S.~C.;
  Waghmare,~U.; Lee,~J.~H. Scale-free ferroelectricity induced by flat phonon
  bands in {HfO$_2$}. \emph{Science} \textbf{2020}, \emph{369},
  1343--1347\relax
\mciteBstWouldAddEndPuncttrue
\mciteSetBstMidEndSepPunct{\mcitedefaultmidpunct}
{\mcitedefaultendpunct}{\mcitedefaultseppunct}\relax
\EndOfBibitem
\bibitem[Luo \latin{et~al.}(2020)Luo, Cheng, Yang, Cao, Ma, Yang, Huang, Wei,
  Zheng, Gong, Yu, Xu, Yuan, Li, Tai, Yu, Shang, Liu, Yu, Ren, Lv, and
  Liu]{Luo20p1391}
Luo,~Q. \latin{et~al.}  A highly {CMOS} compatible hafnia-based ferroelectric
  diode. \emph{Nat. Commun.} \textbf{2020}, \emph{11}, 1391\relax
\mciteBstWouldAddEndPuncttrue
\mciteSetBstMidEndSepPunct{\mcitedefaultmidpunct}
{\mcitedefaultendpunct}{\mcitedefaultseppunct}\relax
\EndOfBibitem
\bibitem[Kim \latin{et~al.}(2021)Kim, Kim, and Lee]{Kim21peabe1341}
Kim,~M.-K.; Kim,~I.-J.; Lee,~J.-S. {CMOS}-compatible ferroelectric {NAND} flash
  memory for high-density, low-power, and high-speed three-dimensional memory.
  \emph{Sci. Adv.} \textbf{2021}, \emph{7}, eabe1341\relax
\mciteBstWouldAddEndPuncttrue
\mciteSetBstMidEndSepPunct{\mcitedefaultmidpunct}
{\mcitedefaultendpunct}{\mcitedefaultseppunct}\relax
\EndOfBibitem
\bibitem[Huan \latin{et~al.}(2014)Huan, Sharma, Rossetti, and
  Ramprasad]{Huan14p064111}
Huan,~T.~D.; Sharma,~V.; Rossetti,~G.~A.; Ramprasad,~R. Pathways Towards
  Ferroelectricity in Hafnia. \emph{Phys. Rev. B} \textbf{2014}, \emph{90},
  064111\relax
\mciteBstWouldAddEndPuncttrue
\mciteSetBstMidEndSepPunct{\mcitedefaultmidpunct}
{\mcitedefaultendpunct}{\mcitedefaultseppunct}\relax
\EndOfBibitem
\bibitem[Sang \latin{et~al.}(2015)Sang, Grimley, Schenk, Schroeder, and
  LeBeau]{Sang15p162905}
Sang,~X.; Grimley,~E.~D.; Schenk,~T.; Schroeder,~U.; LeBeau,~J.~M. On the
  Structural Origins of Ferroelectricity in {HfO}$_2$ Thin Films. \emph{Appl.
  Phys. Lett.} \textbf{2015}, \emph{106}, 162905\relax
\mciteBstWouldAddEndPuncttrue
\mciteSetBstMidEndSepPunct{\mcitedefaultmidpunct}
{\mcitedefaultendpunct}{\mcitedefaultseppunct}\relax
\EndOfBibitem
\bibitem[Talley \latin{et~al.}(2021)Talley, Perkins, Diercks, Brennecka, and
  Zakutayev]{Talley21p1488}
Talley,~K.~R.; Perkins,~C.~L.; Diercks,~D.~R.; Brennecka,~G.~L.; Zakutayev,~A.
  Synthesis of {LaWN$_3$} nitride perovskite with polar symmetry.
  \emph{Science} \textbf{2021}, \emph{374}, 1488--1491\relax
\mciteBstWouldAddEndPuncttrue
\mciteSetBstMidEndSepPunct{\mcitedefaultmidpunct}
{\mcitedefaultendpunct}{\mcitedefaultseppunct}\relax
\EndOfBibitem
\bibitem[Fichtner \latin{et~al.}(2019)Fichtner, Wolff, Lofink, Kienle, and
  Wagner]{Fichtner19p114103}
Fichtner,~S.; Wolff,~N.; Lofink,~F.; Kienle,~L.; Wagner,~B. {AlScN}: A {III-V}
  semiconductor based ferroelectric. \emph{J. Appl. Phys.} \textbf{2019},
  \emph{125}, 114103\relax
\mciteBstWouldAddEndPuncttrue
\mciteSetBstMidEndSepPunct{\mcitedefaultmidpunct}
{\mcitedefaultendpunct}{\mcitedefaultseppunct}\relax
\EndOfBibitem
\bibitem[Hayden \latin{et~al.}(2021)Hayden, Hossain, Xiong, Ferri, Zhu,
  Imperatore, Giebink, Trolier-McKinstry, Dabo, and Maria]{Hayden21p044412}
Hayden,~J.; Hossain,~M.~D.; Xiong,~Y.; Ferri,~K.; Zhu,~W.; Imperatore,~M.~V.;
  Giebink,~N.~C.; Trolier-McKinstry,~S.~E.; Dabo,~I.; Maria,~J.
  Ferroelectricity in boron-substituted aluminum nitride thin films.
  \emph{Phys. Rev. Mater.} \textbf{2021}, \emph{5}, 044412\relax
\mciteBstWouldAddEndPuncttrue
\mciteSetBstMidEndSepPunct{\mcitedefaultmidpunct}
{\mcitedefaultendpunct}{\mcitedefaultseppunct}\relax
\EndOfBibitem
\bibitem[Fang \latin{et~al.}(2017)Fang, Fisher, Kuwabara, Shen, Ogawa,
  Moriwake, Huang, and Duan]{Fang17p014111}
Fang,~Y.-W.; Fisher,~C. A.~J.; Kuwabara,~A.; Shen,~X.-W.; Ogawa,~T.;
  Moriwake,~H.; Huang,~R.; Duan,~C.-G. Lattice dynamics and ferroelectric
  properties of the nitride perovskite ${\mathrm{LaWN}}_{3}$. \emph{Phys. Rev.
  B} \textbf{2017}, \emph{95}, 014111\relax
\mciteBstWouldAddEndPuncttrue
\mciteSetBstMidEndSepPunct{\mcitedefaultmidpunct}
{\mcitedefaultendpunct}{\mcitedefaultseppunct}\relax
\EndOfBibitem
\bibitem[Sun \latin{et~al.}(2019)Sun, Bartel, Arca, Bauers, Matthews,
  Orva{\~{n}}anos, Chen, Toney, Schelhas, Tumas, Tate, Zakutayev, Lany, Holder,
  and Ceder]{Sun19p732}
Sun,~W.; Bartel,~C.~J.; Arca,~E.; Bauers,~S.~R.; Matthews,~B.;
  Orva{\~{n}}anos,~B.; Chen,~B.-R.; Toney,~M.~F.; Schelhas,~L.~T.; Tumas,~W.;
  Tate,~J.; Zakutayev,~A.; Lany,~S.; Holder,~A.~M.; Ceder,~G. A map of the
  inorganic ternary metal nitrides. \emph{Nat. Mater.} \textbf{2019},
  \emph{18}, 732--739\relax
\mciteBstWouldAddEndPuncttrue
\mciteSetBstMidEndSepPunct{\mcitedefaultmidpunct}
{\mcitedefaultendpunct}{\mcitedefaultseppunct}\relax
\EndOfBibitem
\bibitem[Hong(2021)]{Hong21p1445}
Hong,~X. Nitride perovskite becomes polar. \emph{Science} \textbf{2021},
  \emph{374}, 1445--1446\relax
\mciteBstWouldAddEndPuncttrue
\mciteSetBstMidEndSepPunct{\mcitedefaultmidpunct}
{\mcitedefaultendpunct}{\mcitedefaultseppunct}\relax
\EndOfBibitem
\bibitem[Gou \latin{et~al.}(2020)Gou, Zhao, Shi, Harada, and
  Rondinelli]{Gou20p2815}
Gou,~G.; Zhao,~M.; Shi,~J.; Harada,~J.~K.; Rondinelli,~J.~M. Anion Ordered and
  Ferroelectric Ruddlesden--Popper Oxynitride Ca3Nb2N2O5 for
  Visible-Light-Active Photocatalysis. \emph{Chemistry of Materials}
  \textbf{2020}, \emph{32}, 2815--2823\relax
\mciteBstWouldAddEndPuncttrue
\mciteSetBstMidEndSepPunct{\mcitedefaultmidpunct}
{\mcitedefaultendpunct}{\mcitedefaultseppunct}\relax
\EndOfBibitem
\bibitem[Cohen \latin{et~al.}(2021)Cohen, Zhu, Takenaka, and
  Strobel]{Cohen21pJ56}
Cohen,~R.; Zhu,~L.; Takenaka,~H.; Strobel,~T. Prediction of New Ferroelectric
  Clathrate and Polar Oxynitrides. APS March Meeting Abstracts. 2021; pp
  J56--005\relax
\mciteBstWouldAddEndPuncttrue
\mciteSetBstMidEndSepPunct{\mcitedefaultmidpunct}
{\mcitedefaultendpunct}{\mcitedefaultseppunct}\relax
\EndOfBibitem
\bibitem[Palummo \latin{et~al.}(2023)Palummo, Re~Fiorentin, Yamashita,
  Castelli, and Giorgi]{Palummo23p1548}
Palummo,~M.; Re~Fiorentin,~M.; Yamashita,~K.; Castelli,~I.~E.; Giorgi,~G. Study
  of Optoelectronic Features in Polar and Nonpolar Polymorphs of the Oxynitride
  Tin-Based Semiconductor InSnO2N. \emph{The Journal of Physical Chemistry
  Letters} \textbf{2023}, \emph{14}, 1548--1555\relax
\mciteBstWouldAddEndPuncttrue
\mciteSetBstMidEndSepPunct{\mcitedefaultmidpunct}
{\mcitedefaultendpunct}{\mcitedefaultseppunct}\relax
\EndOfBibitem
\bibitem[Yu \latin{et~al.}(2020)Yu, Zhang, Zhang, Wang, Wu, and
  Lee]{Yu20p45023}
Yu,~J.; Zhang,~B.; Zhang,~X.; Wang,~Y.; Wu,~K.; Lee,~M.-H. Finding optimal
  mid-infrared nonlinear optical materials in germanates by first-principles
  high-throughput screening and experimental verification. \emph{ACS Applied
  Materials \& Interfaces} \textbf{2020}, \emph{12}, 45023--45035\relax
\mciteBstWouldAddEndPuncttrue
\mciteSetBstMidEndSepPunct{\mcitedefaultmidpunct}
{\mcitedefaultendpunct}{\mcitedefaultseppunct}\relax
\EndOfBibitem
\bibitem[Bouri and Aschauer(2018)Bouri, and Aschauer]{Bouri18p2771}
Bouri,~M.; Aschauer,~U. Bulk and surface properties of the Ruddlesden--Popper
  oxynitride Sr 2 TaO 3 N. \emph{Physical chemistry chemical physics}
  \textbf{2018}, \emph{20}, 2771--2776\relax
\mciteBstWouldAddEndPuncttrue
\mciteSetBstMidEndSepPunct{\mcitedefaultmidpunct}
{\mcitedefaultendpunct}{\mcitedefaultseppunct}\relax
\EndOfBibitem
\bibitem[Dong \latin{et~al.}(2015)Dong, Fang, Shao, Mulligan, Qiu, Cao, and
  Huang]{Dong15p967}
Dong,~Q.; Fang,~Y.; Shao,~Y.; Mulligan,~P.; Qiu,~J.; Cao,~L.; Huang,~J.
  Electron-hole diffusion lengths $>$ 175 $\mu m$ in solution-grown
  CH$_3$NH$_3$PbI$_3$ single crystals. \emph{Science} \textbf{2015},
  \emph{347}, 967--970\relax
\mciteBstWouldAddEndPuncttrue
\mciteSetBstMidEndSepPunct{\mcitedefaultmidpunct}
{\mcitedefaultendpunct}{\mcitedefaultseppunct}\relax
\EndOfBibitem
\bibitem[Brenner \latin{et~al.}(2015)Brenner, Egger, Rappe, Kronik, Hodes, and
  Cahen]{Brenner15p4754}
Brenner,~T.~M.; Egger,~D.~A.; Rappe,~A.~M.; Kronik,~L.; Hodes,~G.; Cahen,~D.
  Are Mobilities in Hybrid Organic-Inorganic Halide Perovskites Actually
  ``High"? \emph{J. Phys. Chem. Lett.} \textbf{2015}, \emph{6},
  4754--4757\relax
\mciteBstWouldAddEndPuncttrue
\mciteSetBstMidEndSepPunct{\mcitedefaultmidpunct}
{\mcitedefaultendpunct}{\mcitedefaultseppunct}\relax
\EndOfBibitem
\bibitem[Butler \latin{et~al.}(2015)Butler, Frost, and Walsh]{Butler15p838}
Butler,~K.~T.; Frost,~J.~M.; Walsh,~A. Ferroelectric materials for solar energy
  conversion: photoferroics revisited. \emph{Energy Environ. Sci.}
  \textbf{2015}, \emph{8}, 838--848\relax
\mciteBstWouldAddEndPuncttrue
\mciteSetBstMidEndSepPunct{\mcitedefaultmidpunct}
{\mcitedefaultendpunct}{\mcitedefaultseppunct}\relax
\EndOfBibitem
\bibitem[Bennett \latin{et~al.}(2008)Bennett, Grinberg, and
  Rappe]{Bennett08p17409}
Bennett,~J.~W.; Grinberg,~I.; Rappe,~A.~M. New Highly Polar Semiconductor
  Ferroelectrics through d$^{8}$ Cation-O Vacancy Substitution into PbTiO$_3$:
  A Theoretical Study. \emph{J. Am. Chem. Soc.} \textbf{2008}, \emph{130},
  17409--17412\relax
\mciteBstWouldAddEndPuncttrue
\mciteSetBstMidEndSepPunct{\mcitedefaultmidpunct}
{\mcitedefaultendpunct}{\mcitedefaultseppunct}\relax
\EndOfBibitem
\bibitem[Huang \latin{et~al.}(2019)Huang, Taniguchi, and
  Miyasaka]{Huang19p14520}
Huang,~P.-J.; Taniguchi,~K.; Miyasaka,~H. Bulk Photovoltaic Effect in a Pair of
  Chiral-Polar Layered Perovskite-Type Lead Iodides Altered by Chirality of
  Organic Cations. \emph{J. Am. Chem. Soc.} \textbf{2019}, \emph{141},
  14520--14523\relax
\mciteBstWouldAddEndPuncttrue
\mciteSetBstMidEndSepPunct{\mcitedefaultmidpunct}
{\mcitedefaultendpunct}{\mcitedefaultseppunct}\relax
\EndOfBibitem
\bibitem[Shi \latin{et~al.}(2019)Shi, Lu, Song, Chen, Liao, Li, Tang, and
  Xiong]{Shi19p18334}
Shi,~P.-P.; Lu,~S.-Q.; Song,~X.-J.; Chen,~X.-G.; Liao,~W.-Q.; Li,~P.-F.;
  Tang,~Y.-Y.; Xiong,~R.-G. Two-Dimensional Organic-Inorganic Perovskite
  Ferroelectric Semiconductors with Fluorinated Aromatic Spacers. \emph{J. Am.
  Chem. Soc.} \textbf{2019}, \emph{141}, 18334--18340\relax
\mciteBstWouldAddEndPuncttrue
\mciteSetBstMidEndSepPunct{\mcitedefaultmidpunct}
{\mcitedefaultendpunct}{\mcitedefaultseppunct}\relax
\EndOfBibitem
\bibitem[Hitoki \latin{et~al.}(2002)Hitoki, Takata, Kondo, Hara, Kobayashi, and
  Domen]{Hitoki02p1698}
Hitoki,~G.; Takata,~T.; Kondo,~J.~N.; Hara,~M.; Kobayashi,~H.; Domen,~K. An
  oxynitride, {TaON}, as an efficient water oxidation photocatalyst under
  visible light irradiation ($\lambda \le$ 500 nm). \emph{ChemComm}
  \textbf{2002}, 1698--1699\relax
\mciteBstWouldAddEndPuncttrue
\mciteSetBstMidEndSepPunct{\mcitedefaultmidpunct}
{\mcitedefaultendpunct}{\mcitedefaultseppunct}\relax
\EndOfBibitem
\bibitem[Yashima \latin{et~al.}(2007)Yashima, Lee, and Domen]{Yashima07p588}
Yashima,~M.; Lee,~Y.; Domen,~K. Crystal Structure and Electron Density of
  Tantalum Oxynitride, a Visible Light Responsive Photocatalyst. \emph{Chem.
  Mater.} \textbf{2007}, \emph{19}, 588--593\relax
\mciteBstWouldAddEndPuncttrue
\mciteSetBstMidEndSepPunct{\mcitedefaultmidpunct}
{\mcitedefaultendpunct}{\mcitedefaultseppunct}\relax
\EndOfBibitem
\bibitem[Kikuchi \latin{et~al.}(2015)Kikuchi, Kouzaki, Kurabuchi, and
  Hato]{Kikuchi15p711}
Kikuchi,~R.; Kouzaki,~T.; Kurabuchi,~T.; Hato,~K. Characterization of
  baddeleyite-structure {NbON} films deposited by RF reactive sputtering for
  solar hydrogen production devices. \emph{Electrochemistry (Tokyo)}
  \textbf{2015}, \emph{83}, 711--714\relax
\mciteBstWouldAddEndPuncttrue
\mciteSetBstMidEndSepPunct{\mcitedefaultmidpunct}
{\mcitedefaultendpunct}{\mcitedefaultseppunct}\relax
\EndOfBibitem
\bibitem[Giannozzi \latin{et~al.}(2009)Giannozzi, Baroni, Bonini, Calandra,
  Car, Cavazzoni, Ceresoli, Chiarotti, Cococcioni, Dabo, \latin{et~al.}
  others]{Giannozzi09p395502}
Giannozzi,~P.; Baroni,~S.; Bonini,~N.; Calandra,~M.; Car,~R.; Cavazzoni,~C.;
  Ceresoli,~D.; Chiarotti,~G.~L.; Cococcioni,~M.; Dabo,~I., \latin{et~al.}
  {QUANTUM ESPRESSO}: a modular and open-source software project for quantum
  simulations of materials. \emph{J. Phys. Condens. Matter} \textbf{2009},
  \emph{21}, 395502\relax
\mciteBstWouldAddEndPuncttrue
\mciteSetBstMidEndSepPunct{\mcitedefaultmidpunct}
{\mcitedefaultendpunct}{\mcitedefaultseppunct}\relax
\EndOfBibitem
\bibitem[Giannozzi \latin{et~al.}(2017)Giannozzi, Andreussi, Brumme, Bunau,
  Nardelli, Calandra, Car, Cavazzoni, Ceresoli, Cococcioni, \latin{et~al.}
  others]{Giannozzi17p465901}
Giannozzi,~P.; Andreussi,~O.; Brumme,~T.; Bunau,~O.; Nardelli,~M.~B.;
  Calandra,~M.; Car,~R.; Cavazzoni,~C.; Ceresoli,~D.; Cococcioni,~M.,
  \latin{et~al.}  Advanced capabilities for materials modelling with {QUANTUM
  ESPRESSO}. \emph{J. Phys. Condens. Matter} \textbf{2017}, \emph{29},
  465901\relax
\mciteBstWouldAddEndPuncttrue
\mciteSetBstMidEndSepPunct{\mcitedefaultmidpunct}
{\mcitedefaultendpunct}{\mcitedefaultseppunct}\relax
\EndOfBibitem
\bibitem[Garrity \latin{et~al.}(2014)Garrity, Bennett, Rabe, and
  Vanderbilt]{Garrity14p446}
Garrity,~K.~F.; Bennett,~J.~W.; Rabe,~K.~M.; Vanderbilt,~D. Pseudopotentials
  for High-Throughput {DFT} Calculations. \emph{Comput. Mater. Sci.}
  \textbf{2014}, \emph{81}, 446--452\relax
\mciteBstWouldAddEndPuncttrue
\mciteSetBstMidEndSepPunct{\mcitedefaultmidpunct}
{\mcitedefaultendpunct}{\mcitedefaultseppunct}\relax
\EndOfBibitem
\bibitem[Togo and Tanaka(2015)Togo, and Tanaka]{Togo15p1}
Togo,~A.; Tanaka,~I. First principles phonon calculations in materials science.
  \emph{Scr. Mater.} \textbf{2015}, \emph{108}, 1--5\relax
\mciteBstWouldAddEndPuncttrue
\mciteSetBstMidEndSepPunct{\mcitedefaultmidpunct}
{\mcitedefaultendpunct}{\mcitedefaultseppunct}\relax
\EndOfBibitem
\bibitem[Kresse and J(1996)Kresse, and J]{Kresse96p11169}
Kresse,~G.; J,~F. Efficient iterative schemes for ab initio total-energy
  calculations using a plane-wave basis set. \emph{Phys. Rev. B} \textbf{1996},
  \emph{54}, 11169--11186\relax
\mciteBstWouldAddEndPuncttrue
\mciteSetBstMidEndSepPunct{\mcitedefaultmidpunct}
{\mcitedefaultendpunct}{\mcitedefaultseppunct}\relax
\EndOfBibitem
\bibitem[Kresse and J(1996)Kresse, and J]{Kresse96p15}
Kresse,~G.; J,~F. Efficiency of ab-initio total energy calculations for metals
  and semiconductors using a plane-wave basis set. \emph{Comput. Mater. Sci.}
  \textbf{1996}, \emph{6}, 15--50\relax
\mciteBstWouldAddEndPuncttrue
\mciteSetBstMidEndSepPunct{\mcitedefaultmidpunct}
{\mcitedefaultendpunct}{\mcitedefaultseppunct}\relax
\EndOfBibitem
\bibitem[Oganov and Glass(2006)Oganov, and Glass]{Oganov06p244704}
Oganov,~A.~R.; Glass,~C.~W. Crystal Structure Prediction Using Ab Initio
  Evolutionary Techniques: Principles and Applications. \emph{J. Chem. Phys.}
  \textbf{2006}, \emph{124}, 244704\relax
\mciteBstWouldAddEndPuncttrue
\mciteSetBstMidEndSepPunct{\mcitedefaultmidpunct}
{\mcitedefaultendpunct}{\mcitedefaultseppunct}\relax
\EndOfBibitem
\bibitem[Lyakhov \latin{et~al.}(2013)Lyakhov, Oganov, Stokes, and
  Zhu]{Lyakhov13p1172}
Lyakhov,~A.~O.; Oganov,~A.~R.; Stokes,~H.~T.; Zhu,~Q. New Developments in
  Evolutionary Structure Prediction Algorithm {USPEX}. \emph{Comput. Phys.
  Commun.} \textbf{2013}, \emph{184}, 1172--1182\relax
\mciteBstWouldAddEndPuncttrue
\mciteSetBstMidEndSepPunct{\mcitedefaultmidpunct}
{\mcitedefaultendpunct}{\mcitedefaultseppunct}\relax
\EndOfBibitem
\bibitem[Oganov \latin{et~al.}(2011)Oganov, Lyakhov, and Valle]{Oganov11p227}
Oganov,~A.~R.; Lyakhov,~A.~O.; Valle,~M. How Evolutionary Crystal Structure
  Prediction Works{\textemdash}and Why. \emph{Acc. Chem. Res.} \textbf{2011},
  \emph{44}, 227--237\relax
\mciteBstWouldAddEndPuncttrue
\mciteSetBstMidEndSepPunct{\mcitedefaultmidpunct}
{\mcitedefaultendpunct}{\mcitedefaultseppunct}\relax
\EndOfBibitem
\bibitem[Huang \latin{et~al.}(2022)Huang, Hu, and Liu]{Huang22p144106}
Huang,~J.; Hu,~Y.; Liu,~S. Origin of ferroelectricity in magnesium-doped zinc
  oxide. \emph{Phys. Rev. B} \textbf{2022}, \emph{106}, 144106\relax
\mciteBstWouldAddEndPuncttrue
\mciteSetBstMidEndSepPunct{\mcitedefaultmidpunct}
{\mcitedefaultendpunct}{\mcitedefaultseppunct}\relax
\EndOfBibitem
\bibitem[Madsen and Singh(2006)Madsen, and Singh]{Madsen06p67}
Madsen,~G.~K.; Singh,~D.~J. BoltzTraP. A code for calculating band-structure
  dependent quantities. \emph{Comput. Phys. Commun.} \textbf{2006}, \emph{175},
  67--71\relax
\mciteBstWouldAddEndPuncttrue
\mciteSetBstMidEndSepPunct{\mcitedefaultmidpunct}
{\mcitedefaultendpunct}{\mcitedefaultseppunct}\relax
\EndOfBibitem
\bibitem[Liu and Hanrahan(2019)Liu, and Hanrahan]{Liu19p054404}
Liu,~S.; Hanrahan,~B.~M. Effects of growth orientations and epitaxial strains
  on phase stability of {HfO}$_2$ thin films. \emph{Phys. Rev. Mater.}
  \textbf{2019}, \emph{3}, 054404\relax
\mciteBstWouldAddEndPuncttrue
\mciteSetBstMidEndSepPunct{\mcitedefaultmidpunct}
{\mcitedefaultendpunct}{\mcitedefaultseppunct}\relax
\EndOfBibitem
\bibitem[Fan \latin{et~al.}(2022)Fan, Singh, Xu, Park, Qi, Cheong, Vanderbilt,
  Rabe, and Musfeldt]{Fan22p32}
Fan,~S.; Singh,~S.; Xu,~X.; Park,~K.; Qi,~Y.; Cheong,~S.~W.; Vanderbilt,~D.;
  Rabe,~K.~M.; Musfeldt,~J.~L. Vibrational fingerprints of ferroelectric
  {HfO}$_2$. \emph{npj Quantum Mater.} \textbf{2022}, \emph{7}, 32\relax
\mciteBstWouldAddEndPuncttrue
\mciteSetBstMidEndSepPunct{\mcitedefaultmidpunct}
{\mcitedefaultendpunct}{\mcitedefaultseppunct}\relax
\EndOfBibitem
\bibitem[Wei \latin{et~al.}(2022)Wei, Zhao, Zhan, Zhang, Sang, Wang, Tai, Luo,
  Li, Li, and Chen]{Wei22p154101}
Wei,~W.; Zhao,~G.; Zhan,~X.; Zhang,~W.; Sang,~P.; Wang,~Q.; Tai,~L.; Luo,~Q.;
  Li,~Y.; Li,~C.; Chen,~J. Switching pathway-dependent strain-effects on the
  ferroelectric properties and structural deformations in orthorhombic HfO$_2$.
  \emph{J. Appl. Phys.} \textbf{2022}, \emph{131}, 154101\relax
\mciteBstWouldAddEndPuncttrue
\mciteSetBstMidEndSepPunct{\mcitedefaultmidpunct}
{\mcitedefaultendpunct}{\mcitedefaultseppunct}\relax
\EndOfBibitem
\bibitem[Choe \latin{et~al.}(2021)Choe, Kim, Moon, Jo, Bae, Nam, Lee, and
  Heo]{Choe21p8}
Choe,~D.-H.; Kim,~S.; Moon,~T.; Jo,~S.; Bae,~H.; Nam,~S.-G.; Lee,~Y.~S.;
  Heo,~J. Unexpectedly low barrier of ferroelectric switching in HfO$_2$ via
  topological domain walls. \emph{Materials Today} \textbf{2021}, \emph{50},
  8--15\relax
\mciteBstWouldAddEndPuncttrue
\mciteSetBstMidEndSepPunct{\mcitedefaultmidpunct}
{\mcitedefaultendpunct}{\mcitedefaultseppunct}\relax
\EndOfBibitem
\bibitem[King-Smith and Vanderbilt(1993)King-Smith, and
  Vanderbilt]{KingSmith93p1651}
King-Smith,~R.~D.; Vanderbilt,~D. Theory of polarization of crystalline solids.
  \emph{Phys. Rev. B} \textbf{1993}, \emph{47}, 1651\relax
\mciteBstWouldAddEndPuncttrue
\mciteSetBstMidEndSepPunct{\mcitedefaultmidpunct}
{\mcitedefaultendpunct}{\mcitedefaultseppunct}\relax
\EndOfBibitem
\bibitem[Spaldin(2012)]{Spaldin12p2}
Spaldin,~N.~A. A beginner{\textquotesingle}s guide to the modern theory of
  polarization. \emph{J Solid State Chem} \textbf{2012}, \emph{195},
  2--10\relax
\mciteBstWouldAddEndPuncttrue
\mciteSetBstMidEndSepPunct{\mcitedefaultmidpunct}
{\mcitedefaultendpunct}{\mcitedefaultseppunct}\relax
\EndOfBibitem
\bibitem[Zhong \latin{et~al.}(1994)Zhong, King-Smith, and
  Vanderbilt]{Zhong94p3618}
Zhong,~W.; King-Smith,~R.~D.; Vanderbilt,~D. Giant {LO--TO} Splittings in
  Perovskite Ferroelectrics. \emph{Phys. Rev. Lett.} \textbf{1994}, \emph{72},
  3618--22\relax
\mciteBstWouldAddEndPuncttrue
\mciteSetBstMidEndSepPunct{\mcitedefaultmidpunct}
{\mcitedefaultendpunct}{\mcitedefaultseppunct}\relax
\EndOfBibitem
\bibitem[Liu and Cohen(2017)Liu, and Cohen]{Liu17p207601}
Liu,~S.; Cohen,~R.~E. Origin of Negative Longitudinal Piezoelectric Effect.
  \emph{Phys. Rev. Lett.} \textbf{2017}, \emph{119}, 207601\relax
\mciteBstWouldAddEndPuncttrue
\mciteSetBstMidEndSepPunct{\mcitedefaultmidpunct}
{\mcitedefaultendpunct}{\mcitedefaultseppunct}\relax
\EndOfBibitem
\bibitem[S\'aghi-Szab\'o \latin{et~al.}(1998)S\'aghi-Szab\'o, Cohen, and
  Krakauer]{Szabo98p4321}
S\'aghi-Szab\'o,~G.; Cohen,~R.~E.; Krakauer,~H. First-Principles Study of
  Piezoelectricity in ${\mathrm{PbTiO}}_{3}$. \emph{Phys. Rev. Lett.}
  \textbf{1998}, \emph{80}, 4321--4324\relax
\mciteBstWouldAddEndPuncttrue
\mciteSetBstMidEndSepPunct{\mcitedefaultmidpunct}
{\mcitedefaultendpunct}{\mcitedefaultseppunct}\relax
\EndOfBibitem
\bibitem[Frohna \latin{et~al.}(2018)Frohna, Deshpande, Harter, Peng, Barker,
  Neaton, Louie, Bakr, Hsieh, and Bernardi]{Frohna18p1829}
Frohna,~K.; Deshpande,~T.; Harter,~J.; Peng,~W.; Barker,~B.~A.; Neaton,~J.~B.;
  Louie,~S.~G.; Bakr,~O.~M.; Hsieh,~D.; Bernardi,~M. Inversion symmetry and
  bulk Rashba effect in methylammonium lead iodide perovskite single crystals.
  \emph{Nat. Commun.} \textbf{2018}, \emph{9}, 1829\relax
\mciteBstWouldAddEndPuncttrue
\mciteSetBstMidEndSepPunct{\mcitedefaultmidpunct}
{\mcitedefaultendpunct}{\mcitedefaultseppunct}\relax
\EndOfBibitem
\bibitem[Motta \latin{et~al.}(2015)Motta, El-Mellouhi, and
  Sanvito]{Motta15p12746}
Motta,~C.; El-Mellouhi,~F.; Sanvito,~S. Charge carrier mobility in hybrid
  halide perovskites. \emph{Sci. Rep.} \textbf{2015}, \emph{5}, 12746\relax
\mciteBstWouldAddEndPuncttrue
\mciteSetBstMidEndSepPunct{\mcitedefaultmidpunct}
{\mcitedefaultendpunct}{\mcitedefaultseppunct}\relax
\EndOfBibitem
\bibitem[Agapito \latin{et~al.}(2013)Agapito, Ferretti, Calzolari, Curtarolo,
  and Nardelli]{Agapito13p165127}
Agapito,~L.~A.; Ferretti,~A.; Calzolari,~A.; Curtarolo,~S.; Nardelli,~M.~B.
  Effective and accurate representation of extended {Bloch} states on finite
  {Hilbert} spaces. \emph{Phys. Rev. B} \textbf{2013}, \emph{88}, 165127\relax
\mciteBstWouldAddEndPuncttrue
\mciteSetBstMidEndSepPunct{\mcitedefaultmidpunct}
{\mcitedefaultendpunct}{\mcitedefaultseppunct}\relax
\EndOfBibitem
\bibitem[Lee and Son(2020)Lee, and Son]{Lee20p043410}
Lee,~S.-H.; Son,~Y.-W. First-principles approach with a pseudohybrid density
  functional for extended {Hubbard} interactions. \emph{Phys. Rev. Research}
  \textbf{2020}, \emph{2}, 043410\relax
\mciteBstWouldAddEndPuncttrue
\mciteSetBstMidEndSepPunct{\mcitedefaultmidpunct}
{\mcitedefaultendpunct}{\mcitedefaultseppunct}\relax
\EndOfBibitem
\bibitem[Tancogne-Dejean and Rubio(2020)Tancogne-Dejean, and
  Rubio]{Tancogne-Dejean20p155117}
Tancogne-Dejean,~N.; Rubio,~A. Parameter-free hybridlike functional based on an
  extended {Hubbard} model: {{DFT}+$U$+$V$}. \emph{Phys. Rev. B} \textbf{2020},
  \emph{102}, 155117\relax
\mciteBstWouldAddEndPuncttrue
\mciteSetBstMidEndSepPunct{\mcitedefaultmidpunct}
{\mcitedefaultendpunct}{\mcitedefaultseppunct}\relax
\EndOfBibitem
\bibitem[Huang \latin{et~al.}(2020)Huang, Lee, Son, Supka, and
  Liu]{Huang20p165157}
Huang,~J.; Lee,~S.-H.; Son,~Y.-W.; Supka,~A.; Liu,~S. First-principles study of
  two-dimensional ferroelectrics using self-consistent Hubbard parameters.
  \emph{Phys. Rev. B} \textbf{2020}, \emph{102}, 165157\relax
\mciteBstWouldAddEndPuncttrue
\mciteSetBstMidEndSepPunct{\mcitedefaultmidpunct}
{\mcitedefaultendpunct}{\mcitedefaultseppunct}\relax
\EndOfBibitem
\bibitem[Yang \latin{et~al.}(2022)Yang, Zhu, and Liu]{Yang22p195159}
Yang,~J.; Zhu,~T.; Liu,~S. Onsite and intersite electronic correlations in the
  Hubbard model for halide perovskites. \emph{Phys. Rev. B} \textbf{2022},
  \emph{106}, 195159\relax
\mciteBstWouldAddEndPuncttrue
\mciteSetBstMidEndSepPunct{\mcitedefaultmidpunct}
{\mcitedefaultendpunct}{\mcitedefaultseppunct}\relax
\EndOfBibitem
\bibitem[Xiang \latin{et~al.}(2013)Xiang, Huang, Kan, Wei, and
  Gong]{Xiang13p118702}
Xiang,~H.~J.; Huang,~B.; Kan,~E.; Wei,~S.-H.; Gong,~X.~G. Towards Direct-Gap
  Silicon Phases by the Inverse Band Structure Design Approach. \emph{Phys.
  Rev. Lett.} \textbf{2013}, \emph{110}, 118702\relax
\mciteBstWouldAddEndPuncttrue
\mciteSetBstMidEndSepPunct{\mcitedefaultmidpunct}
{\mcitedefaultendpunct}{\mcitedefaultseppunct}\relax
\EndOfBibitem
\bibitem[Grinberg \latin{et~al.}(2013)Grinberg, West, Torres, Gou, Stein, Wu,
  Chen, Gallo, Akbashev, Davies, Spanier, and Rappe]{Grinberg13p509}
Grinberg,~I.; West,~D.~V.; Torres,~M.; Gou,~G.; Stein,~D.~M.; Wu,~L.; Chen,~G.;
  Gallo,~E.~M.; Akbashev,~A.~R.; Davies,~P.~K.; Spanier,~J.~E.; Rappe,~A.~M.
  Perovskites Oxides for Visible-light-adsorbing Ferroelectric and Photovoltaic
  Materials. \emph{Nature} \textbf{2013}, \emph{503}, 509--512\relax
\mciteBstWouldAddEndPuncttrue
\mciteSetBstMidEndSepPunct{\mcitedefaultmidpunct}
{\mcitedefaultendpunct}{\mcitedefaultseppunct}\relax
\EndOfBibitem
\bibitem[Xiao \latin{et~al.}(2018)Xiao, Dong, Guo, Wang, Zhong, Li, and
  Yang]{Xiao18p1805802}
Xiao,~H.; Dong,~W.; Guo,~Y.; Wang,~Y.; Zhong,~H.; Li,~Q.; Yang,~M.-M. Design
  for Highly Piezoelectric and Visible/Near-Infrared Photoresponsive Perovskite
  Oxides. \emph{Adv. Mater.} \textbf{2018}, \emph{31}, 1805802\relax
\mciteBstWouldAddEndPuncttrue
\mciteSetBstMidEndSepPunct{\mcitedefaultmidpunct}
{\mcitedefaultendpunct}{\mcitedefaultseppunct}\relax
\EndOfBibitem
\bibitem[Lima \latin{et~al.}(2010)Lima, Moreira, Cioldin, Diniz, and
  Doi]{Lima10p319}
Lima,~L.; Moreira,~M.~D.; Cioldin,~F.; Diniz,~J.~A.; Doi,~I. Tantalum Nitride
  as Promising Gate Electrode for {MOS} Technology. \emph{ECS Trans}
  \textbf{2010}, \emph{31}, 319\relax
\mciteBstWouldAddEndPuncttrue
\mciteSetBstMidEndSepPunct{\mcitedefaultmidpunct}
{\mcitedefaultendpunct}{\mcitedefaultseppunct}\relax
\EndOfBibitem
\bibitem[Zhu \latin{et~al.}(2013)Zhu, Yuan, Li, Richter, Kirillov, Ioannou, and
  Li]{Zhu13p1151}
Zhu,~H.; Yuan,~H.; Li,~H.; Richter,~C.~A.; Kirillov,~O.; Ioannou,~D.~E.; Li,~Q.
  Design and Fabrication of Ta$_2$O$_5$ Stacks for Discrete Multibit Memory
  Application. \emph{IEEE Trans Nanotechnol} \textbf{2013}, \emph{12},
  1151--1157\relax
\mciteBstWouldAddEndPuncttrue
\mciteSetBstMidEndSepPunct{\mcitedefaultmidpunct}
{\mcitedefaultendpunct}{\mcitedefaultseppunct}\relax
\EndOfBibitem
\bibitem[Matsui \latin{et~al.}(2005)Matsui, Hiratani, Kimura, and
  Asano]{Matsui05pF54}
Matsui,~Y.; Hiratani,~M.; Kimura,~S.; Asano,~I. Combining Ta$_2$O$_5$ and
  Nb$_2$O$_5$ in Bilayered Structures and Solid Solutions for Use in {MIM}
  Capacitors. \emph{J. Electrochem. Soc.} \textbf{2005}, \emph{152}, F54\relax
\mciteBstWouldAddEndPuncttrue
\mciteSetBstMidEndSepPunct{\mcitedefaultmidpunct}
{\mcitedefaultendpunct}{\mcitedefaultseppunct}\relax
\EndOfBibitem
\bibitem[Bower \latin{et~al.}(2020)Bower, Loch, Ware, Berenov, Zou, Hovsepian,
  Ehiasarian, and Petrov]{Bower20p45444}
Bower,~R.; Loch,~D. A.~L.; Ware,~E.; Berenov,~A.; Zou,~B.; Hovsepian,~P.~E.;
  Ehiasarian,~A.~P.; Petrov,~P.~K. Complementary Metal-Oxide-Semiconductor
  Compatible Deposition of Nanoscale Transition-Metal Nitride Thin Films for
  Plasmonic Applications. \emph{ACS Appl. Mater. Interfaces} \textbf{2020},
  \emph{12}, 45444--45452\relax
\mciteBstWouldAddEndPuncttrue
\mciteSetBstMidEndSepPunct{\mcitedefaultmidpunct}
{\mcitedefaultendpunct}{\mcitedefaultseppunct}\relax
\EndOfBibitem
\bibitem[Frost \latin{et~al.}(2014)Frost, Butler, Brivio, Hendon, van
  Schilfgaarde, and Walsh]{Frost14p2584}
Frost,~J.~M.; Butler,~K.~T.; Brivio,~F.; Hendon,~C.~H.; van Schilfgaarde,~M.;
  Walsh,~A. Atomistic Origins of High-Performance in Hybrid Halide Perovskite
  Solar Cells. \emph{Nano Lett.} \textbf{2014}, \emph{14}, 2584--2590\relax
\mciteBstWouldAddEndPuncttrue
\mciteSetBstMidEndSepPunct{\mcitedefaultmidpunct}
{\mcitedefaultendpunct}{\mcitedefaultseppunct}\relax
\EndOfBibitem
\bibitem[Liu \latin{et~al.}(2015)Liu, Zheng, Koocher, Takenaka, Wang, and
  Rappe]{Liu15p693}
Liu,~S.; Zheng,~F.; Koocher,~N.~Z.; Takenaka,~H.; Wang,~F.; Rappe,~A.~M.
  Ferroelectric Domain Wall Induced Band Gap Reduction and Charge Separation in
  Organometal Halide Perovskites. \emph{J. Phys. Chem. Lett.} \textbf{2015},
  \emph{6}, 693--699\relax
\mciteBstWouldAddEndPuncttrue
\mciteSetBstMidEndSepPunct{\mcitedefaultmidpunct}
{\mcitedefaultendpunct}{\mcitedefaultseppunct}\relax
\EndOfBibitem
\bibitem[Weishaupt and Str{\"a}hle(1977)Weishaupt, and
  Str{\"a}hle]{Weishaupt77p261}
Weishaupt,~M.; Str{\"a}hle,~J. Darstellung der Oxidnitride VON, NbON und TaON.
  Die Kristallstruktur von NbON und TaON. \emph{Zeitschrift f{\"u}r
  anorganische und allgemeine Chemie} \textbf{1977}, \emph{429}, 261--269\relax
\mciteBstWouldAddEndPuncttrue
\mciteSetBstMidEndSepPunct{\mcitedefaultmidpunct}
{\mcitedefaultendpunct}{\mcitedefaultseppunct}\relax
\EndOfBibitem
\bibitem[Fang \latin{et~al.}(2001)Fang, Orhan, De~Wijs, Hintzen, De~Groot,
  Marchand, Saillard, \latin{et~al.} others]{Fang01p1248}
Fang,~C.; Orhan,~E.; De~Wijs,~G.; Hintzen,~H.; De~Groot,~R.; Marchand,~R.;
  Saillard,~J.-Y., \latin{et~al.}  The electronic structure of tantalum (oxy)
  nitrides TaON and Ta3N5. \emph{Journal of Materials Chemistry} \textbf{2001},
  \emph{11}, 1248--1252\relax
\mciteBstWouldAddEndPuncttrue
\mciteSetBstMidEndSepPunct{\mcitedefaultmidpunct}
{\mcitedefaultendpunct}{\mcitedefaultseppunct}\relax
\EndOfBibitem
\end{mcitethebibliography}
\end{document}